\documentclass[a4paper,accepted=2021-07-08]{quantumarticle}
\pdfoutput=1
\usepackage[utf8]{inputenc}
\usepackage[english]{babel}
\usepackage[T1]{fontenc}

\usepackage[pdftex]{graphicx}  
\usepackage{xcolor}
\usepackage{dcolumn}
\usepackage{bm}
\usepackage{amsmath,amsthm,amssymb,mathrsfs,mathtools,amsfonts}
\usepackage{verbatim}
\usepackage{epsfig}
\usepackage{color}
\usepackage{xfrac}
\usepackage[numbers,sort&compress]{natbib}
\usepackage[normalem]{ulem}
\usepackage{braket}
\definecolor{lblue} {RGB}{51,71,158}
\usepackage[colorlinks=true,citecolor=blue,linkcolor=blue,urlcolor=lblue]{hyperref}

\def\beq{\begin{equation}}
\def\eeq{\end{equation}}

\begin{document}

\title{Self-organized topological insulator due to cavity-mediated correlated tunneling}

\author{Titas Chanda}
\email{titas.chanda@uj.edu.pl}
\orcid{0000-0002-2306-7895}
\affiliation{Institute of Theoretical Physics, Jagiellonian University in Krak\'ow, \L{}ojasiewicza 11, 30-348 Krak\'ow, Poland}

\author{Rebecca Kraus} 
\affiliation{Theoretical Physics, Saarland University, Campus E2.6, D--66123 Saarbr\"ucken, Germany}

\author{Giovanna Morigi}
\orcid{0000-0002-1946-3684}
\affiliation{Theoretical Physics, Saarland University, Campus E2.6, D--66123 Saarbr\"ucken, Germany}

\author{Jakub Zakrzewski}
\orcid{0000-0003-0998-9460}
\affiliation{Institute of Theoretical Physics, Jagiellonian University in Krak\'ow, \L{}ojasiewicza 11, 30-348 Krak\'ow, Poland}
\affiliation{Mark Kac Complex Systems Research Center, Jagiellonian University in Krakow, \L{}ojasiewicza 11, 30-348 Krak\'ow, Poland}

\begin{abstract}
 Topological materials have potential applications for quantum technologies. Non-interacting topological materials, such as e.g., topological insulators and superconductors, are 
classified by means of fundamental symmetry classes. It is instead only partially understood how interactions affect topological properties. Here, we discuss a model where topology emerges from the quantum interference between single-particle dynamics and global interactions. The system is composed by soft-core bosons  that interact via global correlated hopping in a one-dimensional lattice. The onset of quantum interference leads to spontaneous breaking of the lattice translational symmetry, the corresponding phase resembles nontrivial states of the celebrated Su-Schriefer-Heeger model. Like the fermionic Peierls instability, the emerging quantum phase is a topological insulator and is found at half fillings. 
{Originating from quantum interference, this topological phase is found in ``exact'' density-matrix renormalization group calculations and is entirely absent in the mean-field approach.}
We argue that these dynamics can be realized in existing experimental platforms, such as cavity quantum electrodynamics setups, where the topological features can be revealed in the light emitted by the resonator. 
\end{abstract}

\section*{Introduction}

Manifestation of topology in physics \cite{Klitzing80, Thouless82} created a revolution 
{which is continuing}
{for} almost four decades. 
With the discovery of topological materials, condensed matter physics has gained a new terrain {where quantum phases of matter are} no longer controlled by local order parameters as in paradigmatic Landau theory of phase transitions but rather by the conservation of certain symmetries \cite{Wen17}.  These new phases of matter, so called symmetry-protected topological (SPT) phases, display edge and surface states 
that can be robust against perturbations, making them genuine candidates for quantum technologies \cite{Senthil15}.

To date, a detailed understanding of noninteracting topological materials, such as e.g., topological insulators and superconductors \cite{Qi11}, has been obtained through a successful classification based on fundamental symmetry classes, the so-called ``ten-fold way'' \cite{Schnyder08, Kitaev09, Chiu16}. On the other hand,  interactions are almost unavoidable in many-body systems. It is therefore a central issue to understand whether SPT phases can survive the inter-particle interactions, or {perhaps} whether {interactions themselves might stabilize SPT phases and even give rise to novel topological properties} \cite{Raghu08, Weeks10, Castro11, Grushin13, Dauphin12, Gonzalez18, Gonzalez19, Gonzalez19b}. These questions are at the center of an active and growing area of research \cite{Hohenadler13, Rachel18}. 
Recent works {have} argued that the range of interactions plays a crucial role on the existence of SPT phases. Specifically, in frustrated antiferromagnetic spin-$1$ chain with power-law decaying $1/r^{\alpha}$ interactions, topological phases {are expected to} survive at any positive value of $\alpha$ \cite{Gong16}. {This behavior shares similarities with the topological properties of  {noninteracting} Kitaev $p$-wave superconductors  \cite{Kitaev01} that are robust against long-range tunneling  and pairing {couplings} \cite{Viyuela16, Viyuela18, Alecce17, Jaeger20}.
{It was found that for infinite-range interactions the one-dimensional Kitaev chain can exhibit edge modes for appropriate boundary conditions \cite{Patrick17}. On the other hand, in spin chains and Hubbard models, the infinite-range {interactions} suppress the onset of hidden order at the basis of the Haldane topological insulator \cite{Gong16hald, Sicks20}.}

In this work,
we {report a novel mechanism} where nontrivial topology emerges from the quantum interference between infinite-range interactions and single-particle dynamics. We consider  bosons in a one-dimensional (1D) lattice, where the global interactions have the form of correlated tunneling resulting from the coupling of the bosons with a harmonic oscillator. 
We identify the conditions under which this coupling spontaneously breaks the discrete lattice translational symmetry and leads to the emergence of non-trivial edge states at half filling. The resulting dynamics resembles the one described by the famous Su-Schrieffer-Heeger (SSH) model \cite{Su79, Su80}. {The topology we report shares analogies with the recent studies of symmetry breaking topological insulators \cite{Kourtis14, Mivehvar17, Gonzalez18, Gonzalez19, Gonzalez19b}. Nevertheless, differing from previous realizations, here the interference between quantum fluctuations and global interactions is essential for the onset of the topological phase and cannot be understood in terms of a mean-field model.}

We argue that these dynamics can be realized, for instance, in many-body cavity quantum electrodynamics (CQED) setups \cite{Ritsch13, Landig16,Zupancic19, Cosme18, Georges18, farokh_review}, like the one illustrated in Fig.~\ref{fig:scheme} highlighting the experimental feasibility of our proposal. In the following, we first give a brief description of the model in relation to the experimental setup. Then, we ascertain the phase diagram of the system at half filling, characterize the topological phase, and point out how to experimentally verify its existence.

\section*{Bose-Hubbard model with global correlated tunneling}

\begin{figure}
\centering
\includegraphics[width=\linewidth]{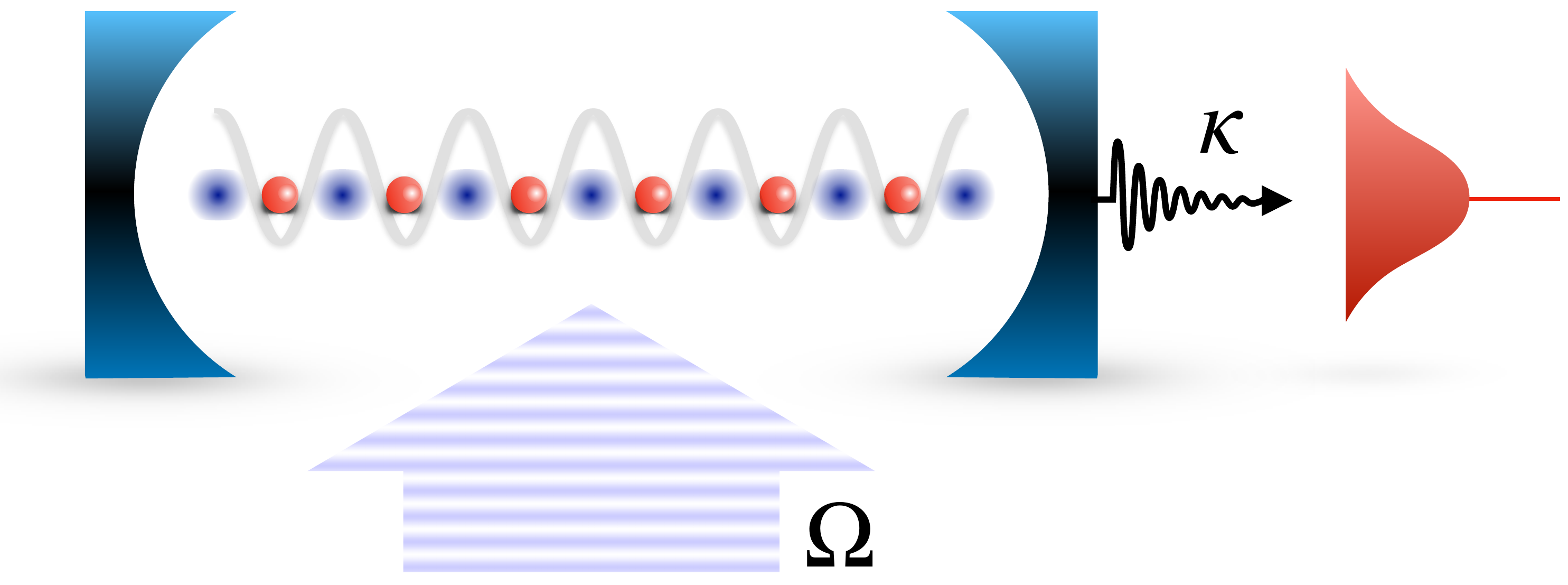}
\caption{{Interference-induced topological phases can be realized in a cavity quantum electrodynamics setup. The bosons are atoms (red spheres) confined by a one-dimensional optical lattice and dispersively interacting with a standing-wave mode of the cavity. The cavity standing-wave field is parallel to the lattice, its wavelength is twice the lattice periodicity, and the atoms are trapped at the nodes of the cavity mode. Correlated hopping originates from coherent scattering of laser light into the cavity, the laser Rabi frequency $\Omega$ controls the strength of the interactions. The field at the cavity output is emitted with rate $\kappa$ and provides information about the phase of the bosons.
}}
\label{fig:scheme}
\end{figure}
The model we consider describes the motion of bosons in a 1D lattice with $N_s$ sites and open boundaries. 
{The specific experimental situation realizing this dynamics is sketched in Fig. \ref{fig:scheme}.} 
The dynamics is governed by the Hamiltonian $\hat H$ in terms of the field operators $\hat b_j$ and $\hat b_j^{\dagger}$, which  annihilate and create, respectively, a boson at site $j$ in the lowest lattice band. 
The bosons, tightly confined in the lowest band of the lattice,  strongly couple with a cavity standing-wave mode (the harmonic oscillator), which is parallel to the lattice. When the atoms are transversely driven by a laser, the Hamiltonian takes the detailed form of a Bose-Hubbard (BH) model with the additional optomechanical coupling with the oscillator \cite{Maschler06}: 
\begin{eqnarray}
\hat H &=& - t \sum_{j=1}^{N_s-1} \left(\hat{b}^\dagger_{j}\hat{b}_{j+1}+{\rm H.c.}\right)+\frac{U}{2}\sum_{j=1}^{N_s}\hat{n}_j\left(\hat{n}_{j}-1\right) \nonumber \\
&&+\,S(\hat a+\hat a^\dagger)y\hat B+\Delta_c \hat a^\dagger \hat a\,,
\label{hammain}
\end{eqnarray}
where we have set $\hbar=1$. Here, the first two terms on the right-hand side (RHS) of Eq. \eqref{hammain} are the nearest-neighbor  hopping with amplitude $t$ and  onsite repulsion with magnitude $U$. The third term on the RHS is the bosons-cavity coupling, where operators $\hat a$ and $\hat a^\dagger$ annihilate and create, respectively, a cavity photon, while the operator $\hat B$ acts on the bosonic Hilbert space. The last term {of \eqref{hammain}} gives the cavity oscillator energy in the reference frame rotating at the frequency of the transverse laser with $\Delta_c$ the detuning between cavity and laser frequency. Finally, the bosons-cavity coupling is scaled by the coefficients $S$  {and $y$ that can be independently tuned. The first one, in fact, is proportional to the amplitude $\Omega$ of the transverse laser field, c.f. Fig. \ref{fig:scheme}. The coefficient $y$, instead, depends on the localization of the scattering atoms at the lattice sites} (see Appendix~\ref{appA}). 

The specific form of operator $\hat B$ depends on the spatial dependence of the cavity-bosons coupling \cite{Habibian13,Caballero15, Caballero16, Elliott16}. {In the present case, we consider} 
\beq
\label{B}
\hat B=\sum_{i=1}^{N_s-1} (-1)^{i+1} \left(\hat{b}^\dagger_{i}\hat{b}_{i+1}+{\rm H.c.} \right).
\eeq
The staggered coupling, $(-1)^{i+1} \left(\hat{b}^\dagger_{i}\hat{b}_{i+1}+{\rm H.c}\right)$, originates from the spatial mode function of the cavity mode along the lattice
when the lattice sites are localized at the nodes of the cavity mode, {and when the periodicity of cavity mode standing wave is twice the periodicity of the optical lattice.}

Hamiltonian \eqref{hammain} {with \eqref{B}} is reminiscent of the phonon-electron coupling {of the SSH model} \cite{Heeger88}. Differing from the ionic lattice of the SSH model, the bosons couple to a single oscillator {-- the cavity mode}. {Here, the} instability is associated with a finite stationary value of the field quadrature $\hat x=\hat a+\hat a^\dagger$: for $\langle \hat x\rangle \neq 0$ the bonds connecting even-odd and odd-even sites differ by a quantity proportional to  $\langle \hat x\rangle$. Since $\hat x$ is a dynamical variable, this process is {accompanied by} a spontaneous breaking of the {lattice translational} symmetry. In a cavity, for instance, $\hat x$ is the electric field that is scattered by the atoms and depends on the atomic mobility. The resulting bosonic dynamics is thus intrinsically nonlinear.

\begin{table*}[t]
\begin{center}
\begin{tabular}{| c || c | c | c | c | c | c | c|}
\hline
Phases & Acronyms &  $\max M_1(k)$ & $k_{\max}$  & $\mathcal{O}_{DW}$ & $\mathcal{O}_B$ & $\mathcal{O}_S$ & $\mathcal{O}_P$ \\
\hline
Superfluid & SF &  $\neq 0$ & $=0$ & $=0$ & $=0$ & $=0$ & $=0$ \\
Bond superfluid & BSF  & $\neq 0$ & $=\pm \pi/2$ & $=0$ & $\neq 0$ & $=0$ & $=0$ \\
Density-wave & DW  & $=0$ & -- & $\neq 0$ & $=0$ & $\neq 0$ & $\neq 0$ \\
Bond insulator & BI  & $=0$ & -- & $=0$ & $\neq 0$ & $\neq 0$ ($=0$) & $=0$ ($\neq 0$) \\
\hline
\end{tabular}
\end{center}
\caption{Different phases, their acronyms, and the corresponding values of order parameters for the  Bose-Hubbard model with cavity-mediated interactions.}
\label{tab:phase}
\end{table*}

We analyze the quantum phases of the system in the limit in which the cavity field (oscillator) can be eliminated from the equations of the bosonic variables assuming that $|\Delta_c|$ is the largest frequency scale of the dynamics. In this limit, the time-averaged field is $\hat \varepsilon (\tau)= \frac{1}{\Delta t}\int_\tau^{\tau+\Delta t} dt \hat a(t) \approx -S\hat B(\tau)/\Delta_c$, where $\tau$ is the coarse-grained time in the grid of step $\Delta t$ \cite{Larson08,Habibian13}. The effective Bose-Hubbard Hamiltonian takes the form
\begin{align}
\hat H =& -t \sum_{j=1}^{N_s-1} \left(\hat{b}^\dagger_{j}\hat{b}_{j+1}+ {\rm H.c.} \right)+\frac{U}{2}\sum_{j=1}^{N_s}\hat{n}_j\left(\hat{n}_{j}-1\right) \nonumber \\
&+\frac{U_1}{N_s}y^2\hat B^2,
\label{ham:1}
\end{align}
where $U_1=S^2N_s/\Delta_c$ and  the explicit dependence on $N_s$ warrants the extensivity of the Hamiltonian in the thermodynamic limit when $U_1$ is fixed by scaling $S\propto 1/\sqrt{N_s}$. The term proportional to $\hat B^2$ emerges from the back-action of the system through the global coupling with the oscillator. It describes a global correlation between pair tunnelings.  

Before discussing the emerging phases, we note that the model in Eq. \eqref{hammain} has extensively been employed for describing ultracold atoms in optical lattices and optomechanically coupled to a cavity mode \cite{Ritsch13,Habibian13,Habibian13b,Caballero16,Landig16}. The specific form of the coupling with operator \eqref{B} has been discussed in Ref. \cite{Caballero16} and can be realized by suitably choosing the sign of the detuning between laser and atomic transition as in Ref. \cite{Zupancic19}. The scaling $1/N_s$ of the nonlinear term corresponds to scaling the quantization volume with the lattice size \cite{Fernandez10}. In the time-scale separation ansatz, the bosons dynamics is coherent provided that $|\Delta_c|$ is also larger than the cavity decay rate $\kappa$: in this regime shot-noise fluctuations are averaged out. Correspondingly, there is no measurement back-action,  {since} the emitted field $\hat \varepsilon (\tau)$ is the statistical average over the time grid $\Delta t$ \cite{Sierant19c}. We note that 
topological phases can also be realized in plethora of platforms where the range of interactions can be tuned and the geometry controlled, prominent examples are optomechanical arrays \cite{Peano15},  photonic systems \cite{Szameit19}, trapped ions \cite{Schneider12}, and ultracold atoms in optical lattices \cite{Ruostekoski02, Javanainen03, Alba11, Tarruell12, Goldman12, Atala13,Ritsch13}.
Specifically,
this model could be also realized in the experimental setups as discussed in \cite{Deng08, Solnyshkov11, Ciuti13}.
In this work, we shall keep on referring to a CQED setup, where the dynamics predicted by Hamiltonian \eqref{hammain} has been extensively studied and can be realized \cite{Ritsch13, Landig16,Zupancic19}.



\section*{Phase diagram at half filling}

In the rest of this work, 
we consider 
{the system at half filling, i.e., }
density $\rho=1/2$. We determine the ground state and its properties using the density matrix renormalization group (DMRG) method \cite{White92, White93} based on matrix product states (MPS) ansatz \cite{Schollwoeck11, Orus14} -- for details see Appendix~\ref{appB}. The phase diagram is determined as a function of $t/U$ and $U_1/U$. The phases and the corresponding observables are summarized in Table \ref{tab:phase}, the observables are detailed below. We characterize off-diagonal long-range order 
by the Fourier transform of the single particle correlations, i.e., the single particle structure factor
\begin{align}
\label{M:1}
&M_1(k)=\frac{1}{N_s^2}\sum_{i,j}e^{ik(i-j)}\left\langle \hat{b}_i^\dagger \hat{b}_j\right\rangle,
\end{align}
which can be experimentally revealed by means of time-of-flight measurements \cite{Greiner02,Kashurnikov02}. The coupling with the cavity induces off-diagonal long-range order, that is signaled by the bond-wave order parameter 
\beq
\mathcal{O}_B= \braket{\hat B}/(2N_s).
\label{O:B}
\eeq 
This quantity is essentially the cavity field in the coarse graining time scale and is directly measurable by heterodyne detection of the electric field emitted by the cavity \cite{Mottl11}. We also consider the density-wave order parameter, $\mathcal{O}_D = \frac{1}{N_s} \left|\braket{\sum_j(-1)^{j} \hat{n}_j} \right|$, which signals the onset of density-wave order and typically characterizes phases when the lattice sites are at the antinodes of the cavity field (see Appendix~\ref{appA}). Moreover, we analyze the behavior of the string and parity order parameters:
\begin{align}
\label{O:S}
&\mathcal{O}_S = 
\braket{\delta \hat{n}_i e^{i \pi \sum_{k=i}^j \delta \hat{n}_k} \delta \hat{n}_j},  
&\mathcal{O}_P = 
\braket{e^{i \pi \sum_{k=i}^j \delta \hat{n}_k}}.
\end{align}
These order parameters depend nonlocally on onsite  fluctuations $\delta \hat{n}_j = \hat{n}_j - \rho$ from the mean density $\rho$. 
{In our calculations of $\mathcal{O}_S$ and $\mathcal{O}_p$, we take $i=10$ and $j = N_s - 11$ to neglect the boundary effects.}

\begin{figure}
\centering
\includegraphics[width=\linewidth]{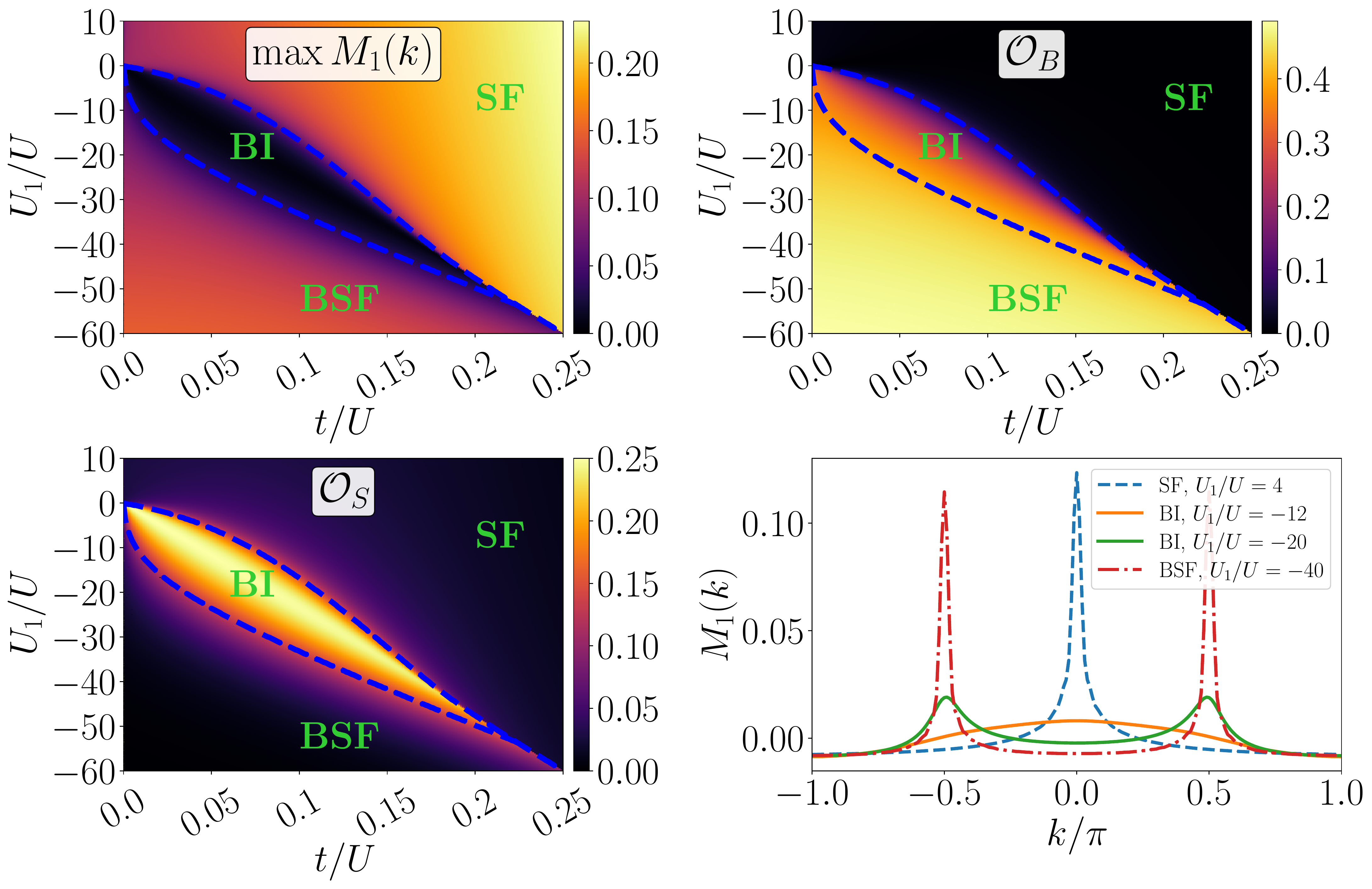}
\caption{{Phase diagram of Hamiltonian \eqref{ham:1} in the $(t/U, U_1/U)$-plane  determined by means of DMRG.} Different panels show the behavior of the {order parameters indicated in the plots. We observe the appearance of an insulating phase  for $t/U \lesssim 0.2$} between standard superfluid (SF) and bond superfluid (BSF) phases {which disappears at larger tunneling values}. The blue dashed lines indicate the borders between the regions where the maximum of $M_1(k)$ evaluated over the ground state crosses a threshold value which we set at $0.03$ to compensate finite-size effects.
The bottom right panel shows the shape of $M_1(k)$ for fixed $t=0.05U$. Sharp peaks appear at $k=0$ for SF and $k=\pm \pi/2$ for BSF in contrast to broad peaks of lower amplitude for BI.  {Note that the effective strength of the correlated hopping in Eq. \eqref{ham:1} scales with $y^2U_1$}. Here, $N_s = 60$ and $y = -0.0658$ (see Appendix~\ref{appA}).}
\label{fig:frog3}
\end{figure}

{We first remark that, for $U_1=0$, hence in the absence of global interactions, the phase is superfluid (SF). We also expect that}
for $U_1>0$ the quantum phase of Hamiltonian in Eq. \eqref{ham:1} is SF, since the formation of a finite cavity field {costs energy}. Instead, we expect that correlated hopping becomes relevant for $U_1<0$.  We have first performed a standard Gutzwiller mean-field  analysis of the model assuming two-site translational invariance. At sufficient large values of $|U_1|$ mean-field predicts the formation of a SF phase of the even or odd bonds accompanied by a finite value of bond-wave order parameter $\mathcal{O}_B$. This phase maximizes the cavity field amplitude and we denote it by Bond Superfluid phase (BSF). We point out that mean-field predicts that the  {ground state} exhibits off-diagonal long-range order for any value of $U_1$. 

{We now discuss the quantum phase obtained from DMRG calculations. The phase diagram is reported as a function of the ratio $t/U$, which scales the strength of the single-particle hopping in units of the onsite repulsion, and of the ratio $U_1/U$, which scales the strength of the correlated hopping.  We {sweep} $U_1$ from positive to negative values. We note that in a cavity the sign of $U_1$ is tuned by means of the sign of the detuning $\Delta_c$. The effective strength, in particular, shall be here scaled by the parameter $y^2$, depending on the particle localization. Here, $y$ is constant across the diagram, since we keep in fact the optical lattice depth constant and tune the ratio $t/U$ by changing the onsite repulsion $U$ {(experimentally, this is realized by means of a Feshbach resonance)}. Figure~\ref{fig:frog3} displays the DMRG results for (a) the maximum value of $|M_1(k)|$ \eqref{M:1}, (b) the bond-wave order parameter \eqref{O:B}, (c) the string order parameter \eqref{O:S}, and (d) the dependence of $M_1(k)$ on the wave number $k$ for different values of $U_1/U$.} 
For $U_1 < 0$ the ground state supports the creation of the cavity field, which is signaled by the finite value of the bond-wave order parameter.  At sufficiently large values of $|U_1/U|$ and $t/U$ the transition is discontinuous, and it separates the SF from the BSF phase, where the effective tunneling amplitudes $\braket{\hat{b}^{\dagger}_i \hat{b}_{i+1} + {\rm H.c.}}$ attain a staggered pattern characterized by a finite value of bond-wave order parameter $\mathcal{O}_B$. The long range coherence of the BSF  phase is manifested by narrow peaks of $M_1(k)$  centered at $k=\pm\pi/2$ (see Fig.~\ref{fig:frog3}). Remarkably, {we observe a reentrant insulating phase separating the SF and the BSF.} The insulator is signaled by vanishing off-diagonal long-range order {and therefore by vanishing structure factor $M_1(k)$}. 
It is characterized by the {non-zero (zero) values of the} string order parameter and by vanishing (non-vanishing) parity order parameter {depending on the boundary sites of these non-local parameters  (see the next section for details)}. 
We denote this phase as a Bond Insulator (BI). This phase is separated from the SF by a continuous phase transition. The transition BI-BSF is also continuous. 

\begin{figure}
\centering
\includegraphics[width=\linewidth]{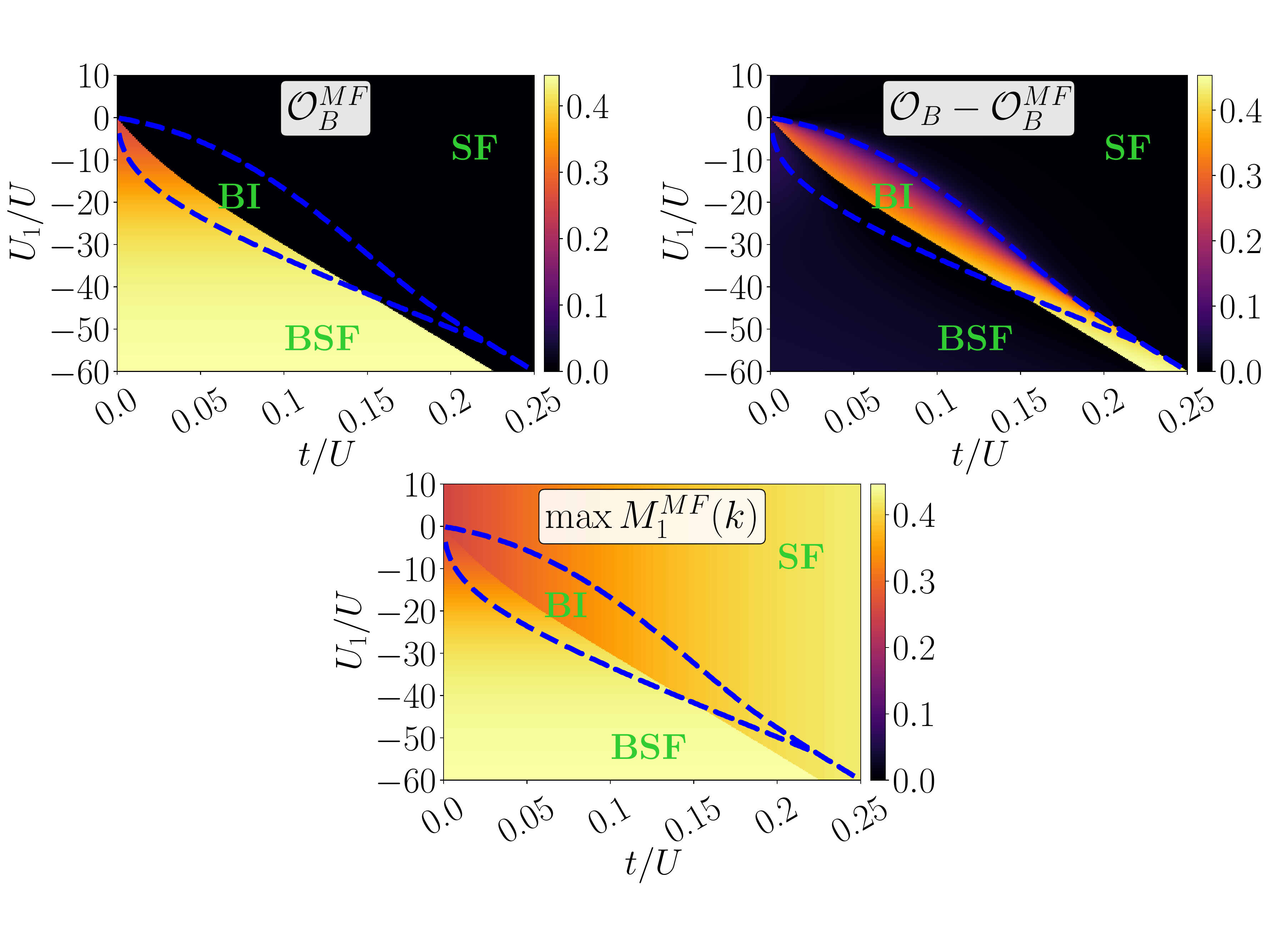}
\caption{Top left panel: Bond-wave order parameter predicted by mean-field Gutzwiller approach, $\mathcal{O}_B^{MF}$, in the $(t/U,U_1/U)$-plane. The top right panel shows the difference ${O}_B-\mathcal{O}_B^{MF}$, where ${O}_B$ has been determined using DMRG. 
{The mean-field Gutzwiller approach for bond-wave order parameter agrees  well with DMRG in SF and BSF regimes, but misses the appearance of BI phase, which is further verified by computing the mean-field results for the maximum of single particle structure factor $\max M_1^{MF}(k)$ (bottom panel), that remains finite in the entire $(t/U,U_1/U)$-plane.}
The blue dashed line indicates the borders between the regions same as in Fig.~\ref{fig:frog3}.}
\label{fig:frogs}
\end{figure}

{We note that the bond insulator phase is entirely absent in {the} standard Gutzwiller mean-field {approach} \cite{Rokhsar91, Krauth92, Sheshadri93,Zwerger03,Zakrzewski05}, {where the bosonic operators are decomposed as 
\begin{equation}
\hat{b}_j = \Phi_j + \delta\hat{b}_j,
\end{equation}  
with $\Phi_j = \braket{\hat{b}_j}$ being the superfluid order parameter in the Gutzwiller analysis. By performing such mean-field analysis with two-site unit cell, we can obtain the bond-wave order parameter  $\mathcal{O}_B^{MF}$ at the mean-field level.}
Figure~\ref{fig:frogs} displays the mean-field bond-wave order  $\mathcal{O}_B^{MF}$  (top left panel) and  the deviation of it from 
the  DMRG result (top right panel).} 
{While the exact borders between various phases quantitatively differ, the mean field Gutzwiller approach agrees well with DMRG in SF and BSF regimes, but misses the appearance of BI phase. To be sure about this, we  check the mean-field value of the maximum of single particle structure factor $M_1(k)$, which is to be redefined as 
\begin{equation}
M_1^{MF}(k)=\frac{1}{N_s^2}\sum_{i,j}e^{ik(i-j)} \Phi_i \Phi_j^*.
\end{equation}
The $\max M_1^{MF}(k)$ is presented in the bottom panel of Fig.~\ref{fig:frogs}. It has possesses high values in the entire  $(t/U,U_1/U)$-plane confirming that the mean-field analysis cannot capture the insulating BI phase where $M_1(k)$ must be vanishingly small.
Moreover, the mean-field value of the string-order parameter $\mathcal{O}_S$ remains exactly zero in the entire phase diagram.} 
{It is to be noted that $\max M_1^{MF}(k)$ does not match the DMRG results presented in Fig.~\ref{fig:frog3}, as in reality the off-diagonal correlations $\braket{\hat{b}_i^\dagger \hat{b}_j}$ decay algebraically with the distance $|i-j|$ in the superfluid phases, while at the mean-field level $\Phi_i \Phi_j^*$ does not.}
Indeed the novel BI phase is entirely due to the interplay between the long-range coupling induced by the cavity and the single-particle tunneling.
{Due to this quantum interference, the insulating BI phase reveals itself in quasi-exact calculations like DMRG, while Gutzwiller analysis can only capture two superfluid phases.}
{We further note that studies of the ground state of \eqref{ham:1}, based on exact diagonalization for small system sizes, did not report the existence of the BI phase \cite{Caballero16}.} In the following section, we characterize the topology associated with the BI phase and argue that it is  stable in the thermodynamic limit.

\section*{Emergent topology associated with the BI phase}

\begin{figure}
\centering
\includegraphics[width=\linewidth]{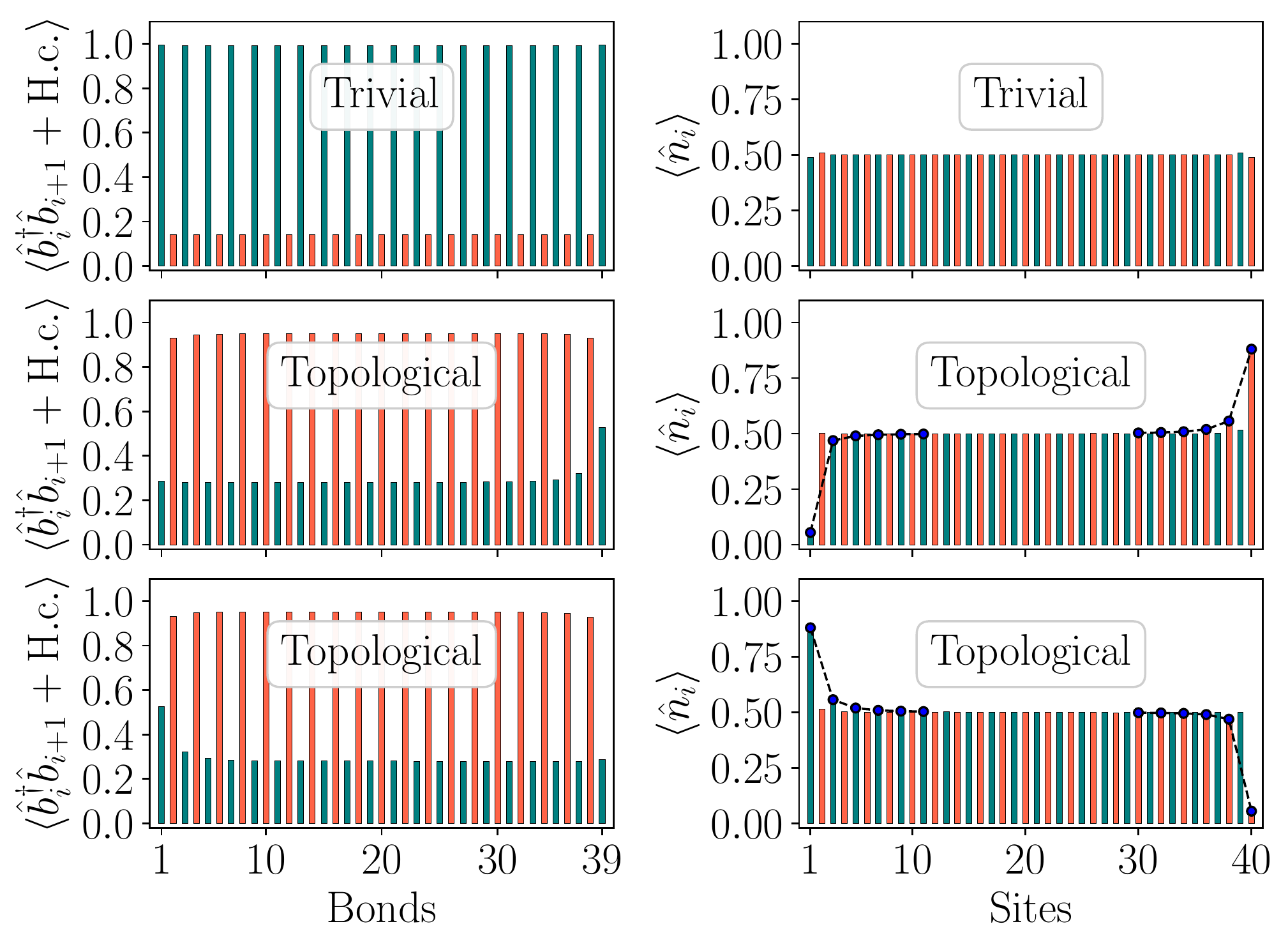}
\caption{{Site-dependent properties of the topological and trivial ground states of the BI phase}. Left panels: Effective tunneling amplitudes $\braket{\hat{b}^{\dagger}_i \hat{b}_{i+1} + {\rm H.c.}}$ {as a function of the bonds $(i,i+1)$. Orange (teal) bars denote the even (odd) bond.}  {Right panels: Density $\braket{\hat n_i}$ as a function of the lattice site. The dashed line is a guide for the eye.} Observe the characteristic alternate weak/strong pattern in the bonds with weak bonds occurring at the edges for topological states that reveal topological particle-hole edge excitations.
{These excitations are due to exactly fractional $\pm 1/2$ particle localizations on the edges.}
 Here, we set $U_1/U = -10$, $t/U = 0.05$, and $y = -0.0658$  (see Appendix~\ref{appA}) to obtain the states using DMRG algorithm.}
\label{fig:pattern}
\end{figure}

By means of excited-state DMRG, we reveal that the BI phase has triply degenerate ground state (quasi-degenerate for finite $N_s$) separated by a finite gap from the other excited states. The site distribution is visualized in Fig.~\ref{fig:pattern} which shows that {the absolute} ground state has a uniform mean half-filling, while the other two states possess edge excitations, namely, fractional $1/2$ particle-hole excitations with respect to the mean half-filling (bottom two rows of Fig.~\ref{fig:pattern}).
Such edge excitations are characterized by the bond-wave order parameter with opposite sign than the trivial phase. They {suggest} that the BI phase is a symmetry protected topological (SPT) phase. 
Similar topological edge states have been reported e.g., for noninteracting system \cite{Atala13} or in superlattice BH model \cite{Grusdt13}, where the superlattice induces a tunneling structure resembling that of the SSH model \cite{Su79, Su80}. 
In our case, instead, the effective tunnelings  {are} spontaneously generated by the creation of a cavity field that breaks {discrete $\mathbb{Z}_2$} translational symmetry of the system. However, bond centered inversion symmetry still remains intact  {-- it} protects this SPT phase. This setting is reminiscent of {a} spontaneous Peierls transition \cite{Peierls01} in bosonic systems, like the ones reported in  \cite{Gonzalez18, Gonzalez19, Gonzalez19b}.



On {a} further inspection, it is found that the string order $\mathcal{O}_S$ and parity order $\mathcal{O}_P$ can be non-zero (zero) depending on the location of the two separated sites that sit at the boundaries of the non-local operators (sites $i$ and $j$ in Eqs.~(\ref{O:S})). We illustrate it in Fig.~\ref{fig:pattern_BI}(a). We find that $\mathcal{O}_S \neq 0$ and $\mathcal{O}_P = 0$, as reported in Fig.~\ref{fig:frog3}, when the non-local operators start at the second site of a strong bond {(i.e., the bond with larger tunneling element)} and end at the first site of a weak bond {(the bond with smaller tunneling element)} further away. 
{That is why we have chosen $i=10$ (even site) and $j=N_s - 11$ (odd site) to calculate these non-local operators for the trivial state in Fig.~\ref{fig:frog3}}.
These are unusual properties when compared to, say, topological Haldane phases of extended BH models at unit filling \cite{Rossini12,Sicks20}.  

\begin{figure}[t]
\centering
\includegraphics[width=\linewidth]{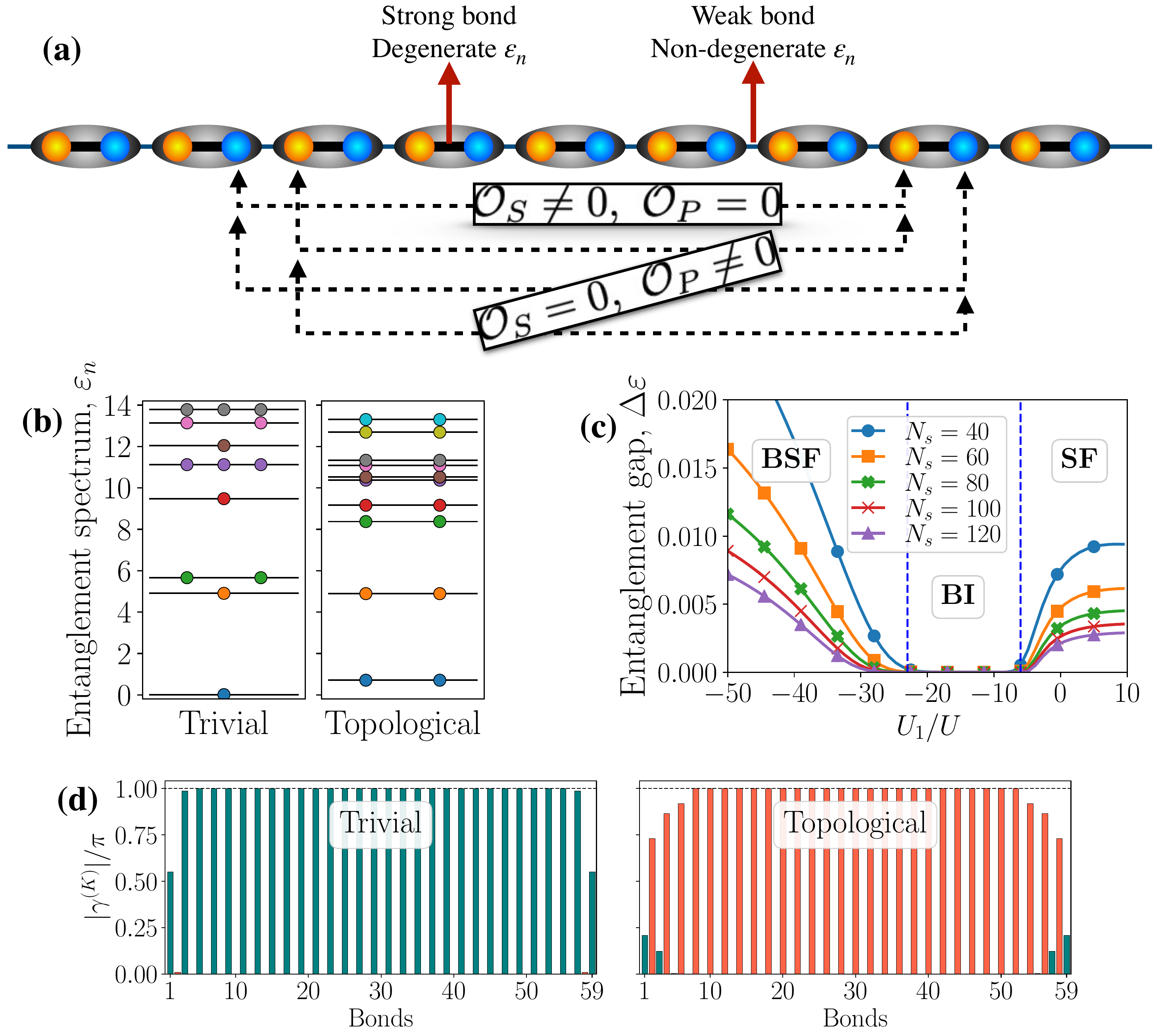}
\caption{(a) Illustration of the BI ground-state properties. The ellipsoids with thick black lines indicate the dimerized strong bonds {with larger effective tunneling amplitudes}, while thin lines indicate the weak alternate bonds {with smaller effective tunneling amplitudes}. The string order $\mathcal{O}_S$ becomes non-zero (with $\mathcal{O}_P = 0$) only when it is measured across blue to orange sites in the figure. In all other cases it vanishes, $\mathcal{O}_S = 0$ and $\mathcal{O}_P \neq 0$. The entanglement spectrum is found to be degenerate in the bulk of the chain when it is measured across the strong bonds, while it is non-degenerate in weak bonds.
(b) Entanglement spectrum $\varepsilon_n$ computed at the middle of the chain of $N_s = 60$ for the trivial and topological states.  For trivial state the spectrum is non-degenerate, while {it is}
doubly degenerate for topological states. (c) Entanglement gap ($\Delta \varepsilon = \varepsilon_1 - \varepsilon_0$) {as a function of $U_1$ and  for fixed $t/U = 0.05$} across BSF-BI-SF phases. {Different curves refer  to different system sizes}.
(d) Local Berry phases $\gamma^{(K)}$ \eqref{Berry}, measured across every bonds for trivial (left) and topological (right) states.
In (b)-(d) other system parameters are the same in Fig. \ref{fig:pattern}.}
\label{fig:pattern_BI}
\end{figure}

To check whether  we are {indeed} dealing with topological states, we calculate the entanglement spectrum of the system. For this purpose we partition the chain into a right ($R$) and left ($L$) subsystem as $\ket{\psi}_{GS} = \sum_n \lambda_n \ket{\psi}_L \otimes \ket{\psi}_R$ where $\lambda_n$ are the corresponding Schmidt coefficients for the specific bipartition. The entanglement spectrum is then defined as the set of all the Schmidt coefficients in logarithmic scale $\varepsilon_n = -2 \log\lambda_n$ and is degenerate for phases with topological properties in one dimension \cite{Pollmann10}. We find that $\varepsilon_n$ are degenerate near the chain center when the bipartition is drawn across a strong bond, while it is non-degenerate at the weak bonds. In Fig.~\ref{fig:pattern_BI}(b), we display the entanglement spectrum for trivial and topological ground states of the BI phase for $N_s=60$, when $\varepsilon_n$ are measured across the bipartition at the chain's center. 
{The entanglement spectrum,} together with the density pattern and the behavior of string and parity order parameters, provide convincing proof of the topological character of the BI phase.
Furthermore, to show that the BI phase is stable in the thermodynamic limit, we consider the entanglement gap, $\Delta \varepsilon = \varepsilon_1 - \varepsilon_0$, for different system sizes. Fig.~\ref{fig:pattern_BI}(c) presents the variation of $\Delta \varepsilon$ across BSF-BI-SF phases for fixed $t/U=0.05$ -- {confirming} the stability of SPT BI phase in the thermodynamic limit.
{We perform similar analysis with different values of $t/U$ too. Such a finite-size analysis with the entanglement gap also confirms that the phase boundaries predicted for $N_s=60$ in Fig.~\ref{fig:frog3} remain stable with increasing system-size.}

In order to reveal the bulk-edge correspondence, we determine the many-body Berry phase \cite{Berry84} and show that it is $\mathbb{Z}_2$-quantized in the BI phase. We first note that, because of the strong interactions, the winding number or the Zak phase \cite{Zak89, Xiao10,Atala13} is not a good topological indicator in  {our case}. Therefore, we follow the original proposition of Hatsugai \cite{Hatsugai06}  and determine the {\it local} many-body Berry phase, which is a topological invariant playing the role of the local ``order parameter'' for {an} interacting case \cite{Hatsugai06}. 
For this purpose, we introduce a local twist $t \rightarrow t e^{i \theta_n}$ in the Hamiltonian \eqref{ham:1}, such that the system still remains gapped in the BI phase. Then the many-body Berry phase is defined as
\begin{equation}
\label{Berry}
\gamma^{(K)} = \text{Arg} \prod_{n=0}^{K-1} \braket{\psi_{\theta_{n+1}}|\psi_{\theta_{n}}},
\end{equation}
where $\psi_{\theta_{n}}$'s are the ground states with $\theta_{0}, \theta_{1}, ..., \theta_{K}= \theta_{0}$ on a loop in $[0, 2 \pi]$. Here, we consider the local Berry phase corresponding to a bond by giving the local twist in tunneling strength $t \rightarrow t e^{i \theta_n}$ only on that particular bond, and take $K = 10$. 
{Since the ground state manifold is (quasi-)degenerate, we need to put small local terms to distinguish between the different states in the manifold in order to compute local many-body Berry phases. Specifically, we add $\mp 0.02\big(\hat{b}^\dagger_{1}\hat{b}_{2} + \hat{b}^\dagger_{N_s-1}\hat{b}_{N_s}  + \text{H.c.}\big)$ respectively for the trivial and topological states to the twisted Hamiltonians. However, in case of topological states, two edge states (bottom two rows of Fig.~\ref{fig:pattern}) are exactly degenerate {\color{blue} and} cannot be distinguished by the local term mentioned above. In that case, to calculate the many-body Berry phase we put one extra particle, i.e., $N_s/2 +1$ bosons in total, so that we have only one edge state where both edges are localized with  {\color{blue} an} extra $+1/2$ particle. Similarly, we could have reduced one particle ($N_s/2-1$ bosons) where the unique edge state would have  {\color{blue} an} extra $-1/2$ particle localization on both the edges.}
The local Berry phases $\gamma^{(K)}$'s are displayed in Fig.~\ref{fig:pattern_BI}(d) for {the} system size $N_s = 60$.  Similar to the entanglement spectrum, we find $\gamma^{(K)} = \pi$ for the strong bonds, while $\gamma^{(K)}=0$ on the weak bonds.

\section*{Discussion}

The BI phase of this model is a  reentrant phase. It separates the SF phase, where correlated hopping is suppressed by quantum fluctuations, from the BSF phase, where correlated tunneling is dominant and single-particle tunneling establishes correlations between the bonds. {We have provided numerical evidence that the emerging topology is essentially characterized by the interplay between quantum fluctuations and correlated tunneling. Interactions are here, therefore, essential for the onset of topology. Their global nature is at the basis of the spontaneous symmetry breaking that accompanies the onset of this phase and which induces an asymmetry between bonds.} {In this respect, it is reminiscent of the Peierls instability of fermions in resonators \cite{Mivehvar17}, where the topology is associated with the spontaneous breaking of $\mathbb{Z}_2$} symmetry. {Differing from that case, where photon scattering gives rise to a self-organized superlattice trapping the atoms, in our model photon scattering interferes with quantum fluctuations.}  Like in \cite{Mivehvar17}, gap and edge states can be measured in the emitted light using pump-probe spectroscopy. The single-particle structure factor may be directly accessible by the time-of-flight momenta distributions \cite{Greiner02,Kashurnikov02} enabling the detection of insulator-superfluid phase transition. The two combined measurements of the cavity output and of the structure factor shall provide a clear distinction between the BI, SF and BSF phases.

{We observe that the global long-range interaction of this model inhibits the formation of solitons.} For other choice of periodicity, and thus of the form of operator $\hat B$, one could expect glassiness associated with the formation of defects \cite{Habibian13}, whose nature is expected to be intrinsically different from the one characterizing short-range interacting structures.

To conclude, we have presented a new paradigm of topological states formation via interference between single particle dynamics and interaction induced hopping.

\begin{acknowledgements}
We thank Daniel Gonz\'alez-Cuadra and Luca Tagliacozzo for discussions on tools and methods, and Shraddha Sharma for helpful comments. 
The support by National Science Centre (Poland) under project Unisono 2017/25/Z/ST2/03029 (T.C.) within QuantERA QTFLAG and OPUS 2019/35/B/ST2/00034 (J.Z.) is acknowledged.
The continuous support of PL-Grid Infrastructure made the reported calculations possible. R.K. and G.M. acknowledge support by the German Research Foundation (the priority program
No. 1929 GiRyd) and by the German Ministry of Education and Research (BMBF) via the QuantERA projects NAQUAS and QTFLAG. Projects NAQUAS and QTFLAG have received funding 
from the QuantERA ERA-NET Cofund in Quantum Technologies implemented within the European Union's Horizon 2020 program.
\end{acknowledgements}

\bibliographystyle{apsrev4-1}
\bibliography{TopoCav.bbl}

\appendix

\section{Coefficients of the extended Bose-Hubbard model}
\label{appA}

To fix the notation, let us consider $N$ atoms of mass $m$ confined within an optical cavity in a quasi-one-dimensional configuration (almost) collinear
with a one-dimensional optical lattice created by light with wave number $k_L=2\pi/\lambda$, which may be different from  $k$ -- the wave number of the cavity field. 
The optical lattice is created by the trapping potential, $V_{\text{trap}} = V_0 \sin^2k_Lx + V (\sin^2k_Ly +  \sin^2k_Lz)$, where $V \gg V_0$ making the atomic motion effectively one-dimensional. In our calculations we take $V = 50E_R$, {where $E_R=\hbar^2 \pi^2/2ma^2$ is the recoil energy and $a$ denotes the periodicity of the optical lattice}. The Wannier basis in the lowest band is $W_i(x,y,z)=w_i(x)\Phi_0(y, z)$ with $w_i(x)$ being the standard one-dimensional Wannier function centered at site $i$ {and} $\Phi_0(y, z)$ - a  two-dimensional Wannier function for {the transverse} deep lattice. {For our purposes, $\Phi_0(y, z)$ is} essentially equivalent to a Gaussian with width $\sigma = \frac{a^2}{\pi^2} \sqrt{E_R/V}$.

After adiabatically eliminating the cavity field one gets an effective Hamiltonian describing atomic motion in terms of
atomic creation/annihilation operators $\hat b_i^\dagger$/$\hat b_i$ for atoms at site $i$ is {decomposed into the sum $\hat H=\hat H_{\rm BH} +\hat H_{\rm Cav}$} \cite{Habibian13,Sierant19c},
where the motion in the lattice and atom-atom interactions are described by the standard Bose-Hubbard Hamiltonian (we assume contact interactions and neglect density-dependent tunneling terms known to be small for such interactions \cite{Dutta15})
\beq
{\hat H_{\rm BH}}= -t \sum_{j=1}^{N_s-1} \left(\hat{b}^\dagger_{j}\hat{b}_{j+1}+{\rm H.c.}\right)+\frac{U}{2}\sum_{j=1}^{N_s}\hat{n}_j\left(\hat{n}_{j}-1\right),
\label{hamBH}
\eeq
where $N_s$ denotes the number of lattice-sites. {Tunneling amplitude $t$ and onsite interaction strength $U$ are given by the integrals}
\begin{align}
t &= \int dx w_i(x)\left(\frac{\hbar^2}{2m} \frac{\partial^2}{\partial x^2}-V_0\sin^2(k_Lx)\right)w_{i+1}(x),  \\                                                                     
U &= g \int d^3r |w_i(x) \Phi_0(y, z)|^4 = \frac{g \pi}{2 a^2} \sqrt{\frac{V}{E_R}} \int dx |w_i(x)|^4.
\end{align}

\begin{figure}
\centering
\includegraphics[width=\linewidth]{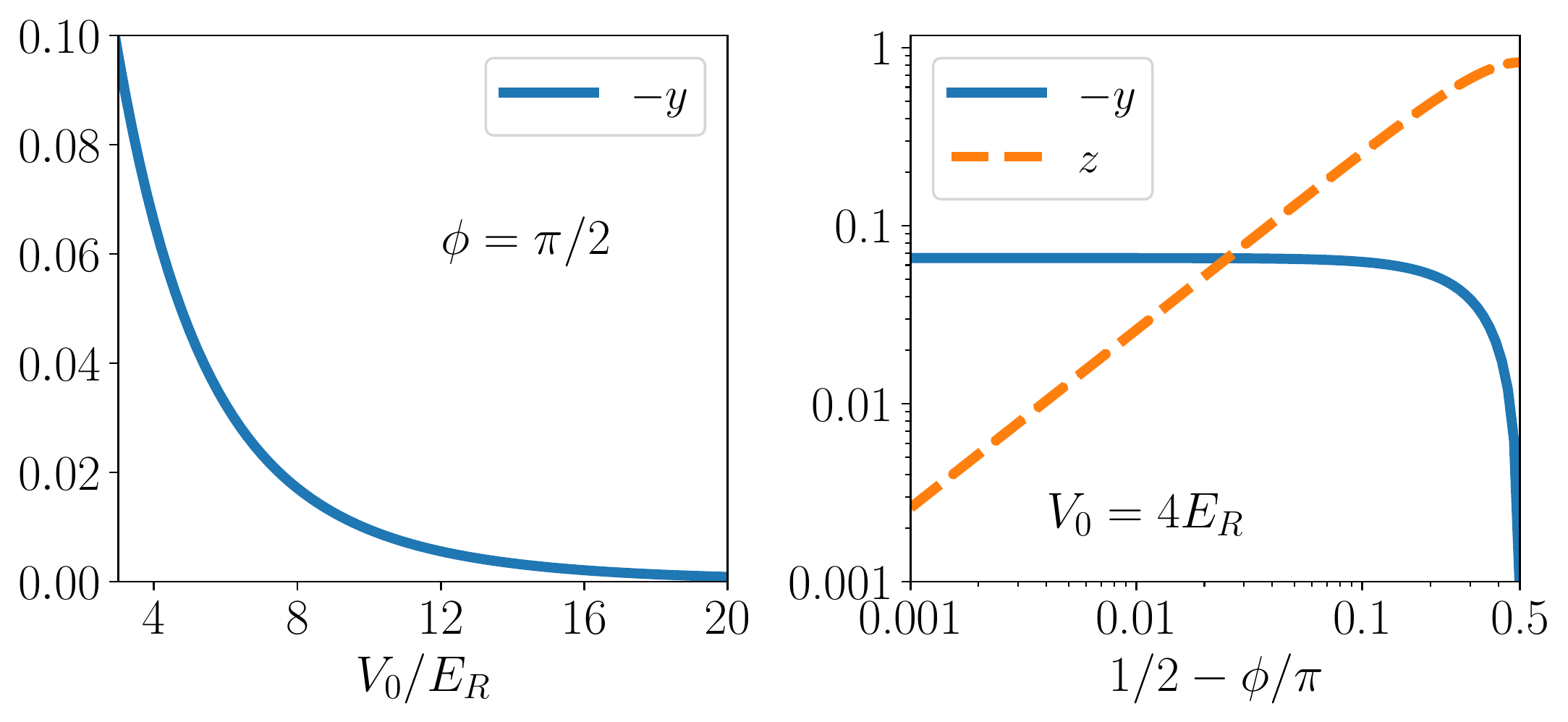}
\caption{(Left) The variation of $y$ as a function of the lattice depth $V_0/E_R$ for $\phi = \pi/2$. Here, $z$ is identically zero for all values of $V_0/E_R$.
(Right) The variations of $y$ and $z$ as $\phi$ deviates from $\pi/2$. We consider $V_0 = 4 E_R$.}
\label{fig:params}
\end{figure}

The rescattering due to the cavity mode leads to a long range ``all-to-all'' interaction terms expressible as
\beq
\hat H_{\rm Cav}=\frac{U_1}{N_s}\left(\sum_{i,j} \hat{b}^\dagger_{i}\hat{b}_{j}\int {\rm d}x\cos(kx+\phi)w_i(x)w_j(x)\right)^2.
\label{eq:hamcav}
\eeq
We rewrite the integrals above as
\begin{equation}
y_{ij}=-\int dx w_i(x)w_{j}(x)\cos(k_L x \beta+\phi).
\label{eq:yij}
\end{equation}
Here, $\beta= k/k_L$ is the ratio between $k$, the wavenumber of the cavity mode, and $k_L$, the wavenumber corresponding to the optical lattice, and $\phi$ is a phase in the mode function. {In our work we consider $\beta=1$, and note that arbitrary values of $\beta$ would lead to a quasiperiodic Hamiltonian \cite{Rojan16,Major18}}.
For $\beta=1$ and
for $i=j$ the {magnitude of the} integral becomes independent of $i$, that can be written as $y_{ii} = (-1)^{i+1} z$ where
\begin{equation}
z = \int dx  w_0^2(x) \cos(k_L x +\phi).
\end{equation}
{Please note here that we have defined $y_{ij}$ (Eq.~\eqref{eq:yij}) coming from Eq.~\eqref{eq:hamcav} with a minus sign to fix $y_{11} = z$ to be positive, as in 
our convention the lattice index starts from $i=1$.}
For non-diagonal $y_{ij}$, we observe that due to localization of Wannier functions $i=j\pm1$ terms may be only significant ones. For our choice of $\beta=1$, the magnitude of the integral again becomes independent of $i$, and can be written as $y_{i,i+1} = y_{i,i-1} = (-1)^{i+1} y$ with
\begin{eqnarray}
y &=& \int dx  w_0(x) w_1(x) \cos(k_L x +\phi), \nonumber \\
 &=& \int dx  w_0(x) w_0(x-a) \cos(k_L x +\phi).
\end{eqnarray}
 Then $\hat H_{\rm Cav}$ may be put in the  form,
\begin{align}
{\hat H_{\rm Cav}}= \frac{U_1}{N_s}\left(z^2 \hat{D}^2 + y z \left(\hat{B}\hat{D} + \hat{D}\hat{B} \right)+ y^2 \hat{B}^2\right)
\end{align}
where
\begin{align}
\hat{D} & =  \sum_{j=1}^{N_s} (-1)^{j+1} \hat{n}_j,  \ \hat{B}  =  \sum_{j=1}^{N_s-1} (-1)^{j+1} \left(\hat{b}_{j}^\dagger \hat{b}_{j+1}+h.c. \right).
\label{opera}
\end{align}

\begin{figure}
\centering
\includegraphics[width=\linewidth]{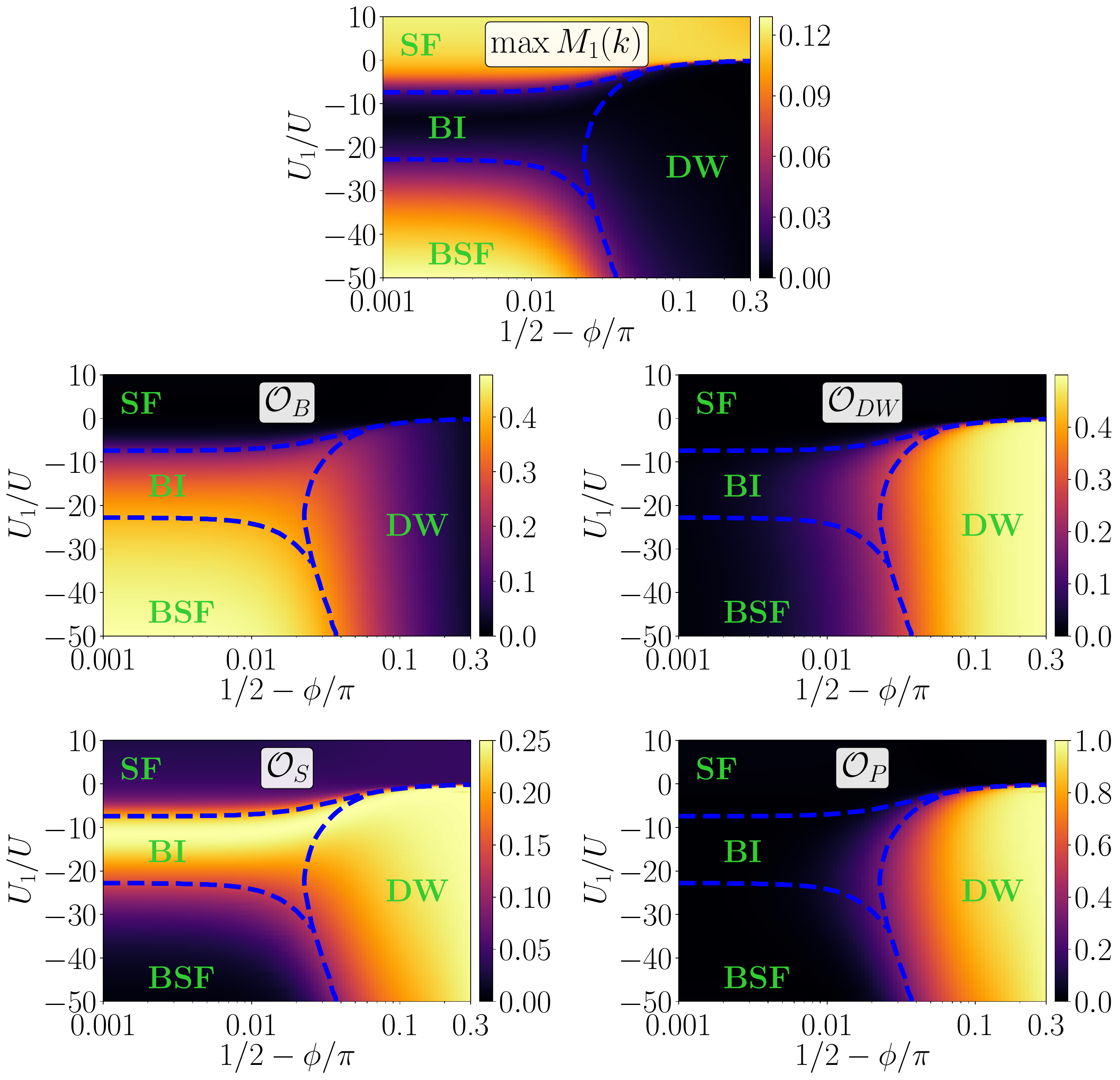}
\caption{Phase diagrams in the $(\phi, U_1/U)$-plane for $t/U = 0.05$ for $N_s = 60$. The BI phase disappears when $\phi$ deviates from $\pi/2$ and density wave phase appears. Here, we choose $V_0$ to be $4 E_R$ so that the values of $y$ and $z$ matches that of Fig. \ref{fig:params} (right panel). Blue dashed lines are guide to the eyes to differentiate different phases.
}
\label{fig:frog4}
\end{figure}

Typically,
one assumes no phase difference ($\phi=0$) between the optical cavity and the external optical lattice, i.e., {the lattice sites are located at the antinodes of the cavity mode}. Then $|z| \neq 0$ and $|y| \approx 0$,
and the terms proportional to $y$ are dropped off leading to the standard case considered in the past \cite{Maschler08}. 
The importance of additional terms was noticed already in \cite{Caballero16} where an identical Hamiltonian is obtained for a {slightly} different arrangement of the cavity and external optical lattice. Here, we have {focused} on a immediate vicinity of $\phi=\pi/2$. {This corresponds to the}  atoms {being} trapped at the nodes of the cavity mode, where $z$ vanishes (see left panel of Fig. \ref{fig:params}). 
However, as $\phi$ starts to deviate from $\pi/2$ the $y$ term rapidly decreases and $z$ term becomes significant (see right panel of Fig. \ref{fig:params}).
Note that the quadratic form of ${\hat H_{\rm Cav}}$ is responsible for the long-range character of the couplings. Squaring $\hat D$
leads then to all-to-all density-density interactions, responsible for a spontaneous formation of density wave phase for sufficient $U_1$ \cite{Dogra16}. 
For the case considered by us, $z \approx 0$ and $\hat B^2$ term leads to the all-to-all long-range correlated tunnelings alternating in sign.
{Throughout the paper, we have fixed $V = 4 E_R$ resulting in $y=-0.0658$ and $z = 0$ for $\phi = \pi/2$.}

In Fig.~\ref{fig:frog4}, we plot the order parameters and the phase diagram in the vicinity of $\phi = \pi/2$ for $t/U = 0.05$. As $\phi$ starts to deviate from $\pi/2$, $y$ starts to diminish and $z$ becomes increasingly larger (see right panel of Fig. \ref{fig:params}). As a result, BI phase is replaced by a more standard  density wave (DW) phase  \cite{Dogra16}, when $\phi$  becomes sufficiently different from $\pi/2$. In the DW phase $\mathcal{O}_{DW}$ as well as $\mathcal{O}_S$ and $\mathcal{O}_P$ are both non-zero, while the structure factor vanishes.

\section{Numerical implementation}
\label{appB}

We use standard matrix product states (MPS) \cite{Schollwoeck11, Orus14} based density matrix renormalization group (DMRG) \cite{White92, White93} method to find the ground state and low-lying excited states of the system with open boundary condition,
where we employ the global $U(1)$ symmetry   corresponding to the conservation of the total number of particles. For that purpose, we use ITensor C++ library (\url{https://itensor.org})
where  the MPO for the all connected long-range Hamiltonian can be constructed exactly \cite{Crosswhite08, Pirvu10} using AutoMPO class.
The maximum number bosons ($n_0$) per site has been truncated to 6, which is justified as we only consider average density to be $\rho = 1/2$. 

We consider random entangled states, $\ket{\psi_{\text{ini}}} = \frac{1}{\sqrt{50}} \sum_{i=0}^{49} \ket{\psi^{\text{rand}}_{i}}$, where $\ket{\psi^{\text{rand}}_{i}}$ are random product states with density $\rho=1/2$, as our initial states for DMRG algorithm. The maximum bond dimension of MPS during standard two-site DMRG sweeps has been restricted to $\chi_{\max} = 200$. We verify the convergence of the DMRG algorithm by checking the deviations in energy in successive DMRG sweeps. When the energy deviation falls below $10^{-12}$, we conclude that the resulting MPS is the ground state of the system.

To obtain low-lying excited states, we first shift the Hamiltonian by a suitable weight factor multiplied with the projector on the previously found state. 
To be precise, for finding the $n^{th}$ excited state $\ket{\psi_n}$, we search for the ground state of the shifted Hamiltonian, 
\begin{equation}
\hat{H}' = \hat{H} + W \sum_{m=0}^{n-1} \ket{\psi_m}\bra{\psi_m},
\end{equation}
 where $W$ should be guessed to be sufficiently larger than $E_n - E_0$.



\end{document}